\documentclass[prl, twocolumn, floatfix, amsmath,superscriptaddress]{revtex4}
\usepackage{graphicx}
\usepackage{amsmath}
\usepackage{amssymb}
\usepackage{amsthm}
\usepackage{amsfonts}
\usepackage[english]{babel}
\usepackage{url}
\usepackage{epstopdf}

\begin{document}
\title{Scattering Lens Resolves sub-$100$~nm Structures with Visible Light}
\author{E.G. van Putten}
\affiliation{Complex Photonic Systems, Faculty of Science and Technology
 and MESA+ Institute for\\ Nanotechnology, University of Twente,
 P.O. Box 217, 7500 AE Enschede, The Netherlands}
\author{D. Akbulut}
\affiliation{Complex Photonic Systems, Faculty of Science and Technology
 and MESA+ Institute for\\ Nanotechnology, University of Twente,
 P.O. Box 217, 7500 AE Enschede, The Netherlands}
 \author{J. Bertolotti}
\affiliation{University of Florence, Dipartimento di Fisica, 50019 Sesto Fiorentino, Italy}
\affiliation{Complex Photonic Systems, Faculty of Science and Technology
 and MESA+ Institute for\\ Nanotechnology, University of Twente,
 P.O. Box 217, 7500 AE Enschede, The Netherlands}
\author{W.L. Vos}
\affiliation{Complex Photonic Systems, Faculty of Science and Technology
 and MESA+ Institute for\\ Nanotechnology, University of Twente,
 P.O. Box 217, 7500 AE Enschede, The Netherlands}
 \author{A. Lagendijk}
\affiliation{Complex Photonic Systems, Faculty of Science and Technology
 and MESA+ Institute for\\ Nanotechnology, University of Twente,
 P.O. Box 217, 7500 AE Enschede, The Netherlands}
 \affiliation{FOM Institute for Atomic and Molecular Physics (AMOLF), Science Park 104, 1098 XG Amsterdam, The Netherlands}
\author{A.P. Mosk}
\affiliation{Complex Photonic Systems, Faculty of Science and Technology
 and MESA+ Institute for\\ Nanotechnology, University of Twente,
 P.O. Box 217, 7500 AE Enschede, The Netherlands}

\date{\today}

\begin{abstract}
The smallest structures that conventional lenses are able to optically resolve
are of the order of $200$~nm.
We introduce a new type of lens that exploits multiple scattering of light to
generate a scanning nano-sized optical focus. With an experimental realization of this lens in
gallium phosphide we have succeeded to image gold nanoparticles at $97$~nm optical resolution.
Our work is the first lens that provides a resolution in the nanometer regime at
visible wavelengths.
\end{abstract}\maketitle


Many essential structures in nanoscience and nanotechnology,
such as cellular organelles, nanoelectronic circuits, and photonic structures,
have spatial features in the order of $100$~nm. The optical resolution of conventional
lenses is limited to approximately $200$~nm
by their numerical aperture and therefore they cannot resolve nanostructure.
With fluorescence based imaging methods it is possible to reconstruct an image of objects
that are a substantial factor smaller than the focus size by exploiting the photophysics of
extrinsic fluorophores.\cite{Hell1994aa, Dyba2002aa, Betzig2006aa, Rust2006aa, Hell2007aa}
Their resolution strongly depends on the shape of the optical focus, which is determined by
conventional lens systems. This dependence makes them vulnerable to focal distortion by
scattering. Moreover, its not always feasible or desirable to dope the object under study.
Other imaging methods improve their resolution by reconstructing the evanescent
waves that decay exponentially with distance from the object. Intricate near
field microscopes bring fragile nano-sized probes in close proximity
of the object where the evanescent field is still measurable.\cite{Pohl1993aa}
With this technique it is hard to quantify the interaction between the short-lived
tip and the structure.
Metamaterials, which are meticulously nanostructured artificial composites, can
be engineered to access the evanescent waves and
image sub-wavelength structures\cite{Pendry2000aa} as demonstrated with
superlenses\cite{Fang2005aa} and hyperlenses\cite{Liu2007aa} in the UV.
These materials physically decrease the focus size, which brings the possibility for
improvement of both linear and non-linear imaging techniques.
In the especially relevant visible range of the spectrum, plasmonic metamaterials
can be used to produce nano-sized isolated hot spots\cite{Stockman2002aa,Aeschlimann2007aa,Kao2011aa}
but the limited control over their position makes them unsuitable for imaging.
Up to now, a freely scannable nano-sized optical focus has not been demonstrated.

\begin{figure}
  \centering
  \includegraphics[width=8.4 cm]{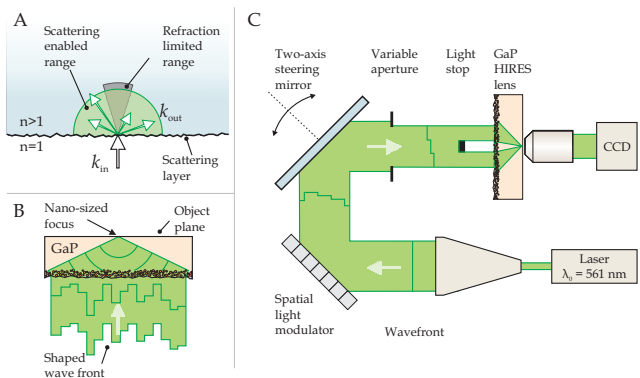}
  \caption{\textbf{(A)} Principle of light coupling to high transversal k-vectors into
  a high-index material. Without scattering refraction would strongly
  limit the angular range to which light could be coupled. By exploiting strong
  scattering at the interface, incident light $k_\text{in}$ is coupled
  to all outgoing angles $k_\text{out}$ in the high index material.
  \textbf{(B)} Schematic of a HIRES-lens that uses light scattering
  to achieve a high optical resolution. This HIRES-lens consists of a slab of gallium phosphide (GaP)
  on top of a strongly scattering porous layer. By controlling the incident wavefront, a
  small focus is made in the object plane of the HIRES-lens.
  \textbf{(C)} Overview of the setup. A $\lambda_0=561$~nm laser beam is expanded and illuminates
  a phase only spatial light modulator. The modulated reflected beam is first imaged onto
  a two-axis steering mirror and then onto the porous surface of the GaP HIRES-lens.
  A variable aperture controls the extent of the illuminated area and a light stop places the
  setup in a dark field configuration by blocking the center of the light beam. We image the
  object plane onto a CCD camera using an oil immersion microscope objective.
  }\label{fig:Fig1}
\end{figure}

In this Letter we introduce a new type of lens that generates a scanning
nano-sized optical focus. We used this lens
to image a collection of gold nanoparticles at 97~nm optical resolution. The lens exploits
multiple scattering of light in a porous high refractive index material to increase the numerical
aperture of the system; a principle we name High Index Resolution Enhancement by Scattering (HIRES).

A HIRES-lens consists of a homogenous slab of high-index material
on top of a strongly disordered scattering layer.
The disordered layer breaks the translational invariance of the interface, which
enables incident light to be coupled to all propagating angles inside the high-refractive-index
material as is shown in Fig.~\ref{fig:Fig1}A.
Yet multiple scattering also scrambles the wavefront creating a speckle-like pattern on the
object plane that itself cannot be used for imaging. Therefore we manipulate the incident
wavefront in order to force constructive interference of the
scattered light at a position in the object plane of our HIRES-lens. The wavefront
is controlled using a feedback based method\cite{Vellekoop2007aa} that is
conceptionally related to phase conjugation\cite{Leith1966aa} and time reversal\cite{Fink2000}.
As a result, a perfectly spherical wave emerges from the porous layer and
converges towards the object plane to form a sharp optical focus (Fig.~\ref{fig:Fig1}B).
Whereas in conventional optics (e.g. solid immersion lenses\cite{Wu1999aa} or total
internal reflection microscopes\cite{Axelrod1984aa}) any inevitable surface roughness causes a
distortion of the wavefront and a concomitant loss of resolution,
the inherent random nature makes a HIRES-lens robust
for these abberations. Any wavefront error is distributed randomly over all
outgoing directions, slightly reducing the contrast but not the
resolution\cite{Vellekoop2010aa}. In order to use the HIRES-lens for high
resolution imaging, the focus is easily moved around in the object
plane by steering the incident wavefront directly exploiting the angular
correlations in the scattered light; an effect well known as the
optical memory effect.\cite{Feng1988aa,Freund1988aa,Vellekoop2010ab}
By raster scanning the focus across an object we
acquire an abberation-free high resolution image. The robust scanning high resolution
focus makes the HIRES-lens excellently suited for optical imaging of nanostructures.

To demonstrate an experimental implementation of our HIRES-lens we fabricate it in
gallium phosphide (GaP). GaP is transparent in a large part of the visible
spectrum ($\lambda_0>550$~nm) and has a maximum refractive index of
n=$3.41$, higher than any other transparent material in this wavelength range.\cite{Aspnes1983aa}
Electrochemically etching GaP with sulfuric acid
(H$_2$SO$_4$) creates macroporous networks resulting in
one of the strongest scattering photonic structures ever
observed.\cite{Schuurmans1999aa} Using this etching process we create a
$d=2.8~\mu$m thick porous layer on one side of a crystalline GaP wafer.
This layer is thick enough to completely randomize the incident wavefront and
to suppress any unscattered background light.

The optical memory effect allows us to shift the scattered light
in the object plane of the HIRES-lens over a distance
$r \approx 1.8 L \lambda /(2 \pi n^2 d)$ before the intensity correlation
decreases to $1/e$\cite{Feng1988aa}, where $L=400~\mu$m is the thickness
of the wafer. The loss of correlation
only affects the intensity in the focus (not its
shape) making it easy to correct for this effect without losing
resolution. Due to the high refractive index contrast on the flat
GaP-air interface, a large fraction of the light is internally reflected.
The reflected light interferes with the light that comes directly from the
porous layer. This interference causes a background signal that is
$3$ times larger than the focus intensity. We have therefore
strongly suppressed the internal reflections by depositing an
approximately $200$~nm thick anti-internal-reflection coating of amorphous silicon
on the surface.
The amorphous silicon is nearly index matched with the GaP
and strongly absorbs the light that would otherwise be internally reflected.
As a result of this layer, the background signal is significantly reduced to
only $0.04$ times the focus intensity\cite{SupportingMaterial_vanPutten2011aa}. The resulting
field of view of our coated HIRES-lens is measured
to be
$r = 1.7 \pm 0.1~\mu$m in radius; $85\%$
of the theoretical limit determined by the
optical memory effect. In the center of the surface we created a
small window of about $10~\mu$m in diameter by locally
removing the anti-internal-reflection coating. We use this window to place objects
onto our HIRES-lens. As a test sample we have deposited a
random configuration of gold nanoparticles with a specified diameter of $50$~nm
inside this window.

An overview of our setup is shown in Fig.~\ref{fig:Fig1}C. We use a CW laser with a
wavelength of $\lambda_0 = 561$~nm just below the GaP bandgap of $2.24$~eV ($550$~nm) where
the refractive index is maximal and the absorption is still negligible.\cite{Aspnes1983aa}
We spatially partition the
wavefront into square segments of which we independently control the phase using a
spatial light modulator (SLM).
The SLM is first imaged onto a two-axis fast steering mirror and then
onto the porous surface of the HIRES-lens. With a variable
aperture we set the radius $R_\text{max}$ of the illuminated surface area between
$0~\mu$m and $400~\mu$m. The visibility of the gold nanoparticles is maximized by
blocking the central part of the illumination ($R~<~196~\mu$m), placing the system
in a dark field configuration. At the back of the HIRES-lens a high-quality oil immersion
microscope objective (NA~=~$1.49$) images the object plane onto
a CCD camera.
This objective is used to efficiently collect all the light scattered from the object plane
and to obtain an reference image which is used as a comparison for our HIRES-lens.
Notice that in our scheme the resolution is determined by the HIRES-lens itself and
does not depend on the imaging optics at the back.

We first synthesize the wavefront that, after being scattered,
creates a focus in the object plane. We use light scattered
from one of the gold nanoparticles in the object plane as a feedback signal to obtain
a set of complex amplitudes that describe the propagation from different incident
positions on the porous layer towards the nanoparticle\cite{Vellekoop2007aa}.
By reversing the phase of
these complex amplitudes we force the light waves to interfere constructively at
the exact location of the nanoparticle.
The focus is moved around in the image plane
by rotating every contributing k-vector over a corresponding angle. We
apply these rotations by adding a deterministic phase pattern to the
incident wavefront. In the paraxial limit, a simple tilt
of the wavefront would suffice to displace the focus.\cite{Vellekoop2010ab,Hsieh2010aa}
For our high resolution focus, which lies beyond this limit, an additional
position dependent phase correction is required that we apply
using the SLM.\cite{SupportingMaterial_vanPutten2011aa} The addition of this
correction is essential for a proper displacement of the focus.

\begin{figure}
  \centering
  \includegraphics[width=8.4 cm]{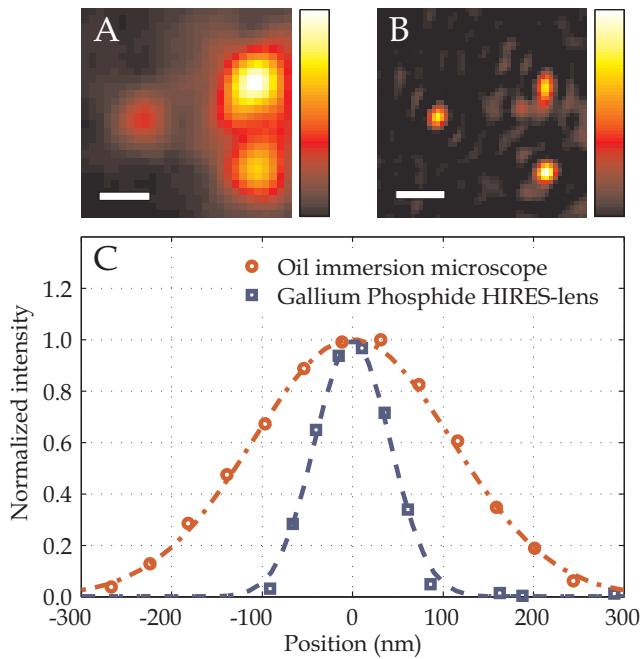}
  \caption{Experimental imaging demonstration with a GaP HIRES-lens.
  \textbf{(A)} A reference image taken with conventional oil immersion microscope
  ($\text{NA} = 1.49$). The image shows a blurred collection of gold nanoparticles. The scale bar
  represents $300$~nm.
  \textbf{(B)} A high resolution image acquired with our GaP HIRES-lens.
  The image was obtained by scanning a small focus
  over the objects while monitoring the amount of scattered light and
  deconvoluted with Eq.~\ref{eq:intensityProfile}\cite{SupportingMaterial_vanPutten2011aa}.
  \textbf{(C)} A vertical cross section through the center of the left sphere
  in \textbf{A} and \textbf{B} shows the increase in
  resolution. The dashed lines are Gaussian fits to the data points.
  }\label{fig:Fig2}
\end{figure}
In Fig.~\ref{fig:Fig2} we show the imaging capabilities of
the GaP HIRES-lens.
First a reference image was acquired with
the high-quality microscope behind the HIRES-lens
(Fig.~\ref{fig:Fig2}A). Because the size of the gold nanoparticles is much
smaller than the resolution limit of this conventional oil immersion
microscope the image of the nanoparticles is blurred.
Next we used our HIRES-lens to construct a high-resolution image. By manipulating
the wavefront a focus was generated on the leftmost nanoparticle. We raster
scanned the focus across the object plane while we constantly monitored
the amount of scattered light. In Fig.~\ref{fig:Fig2}B the result of the scan
is shown\cite{SupportingMaterial_vanPutten2011aa}.
A cross section through the center of the left sphere (Fig.~\ref{fig:Fig2}C)
clearly shows the improvement in resolution we
obtained with our HIRES-lens, confirming our expectations
that the resolution of this image is far better than that of
the conventional high-quality detection optics.

\begin{figure}
  \centering
  \includegraphics[width=8.4 cm]{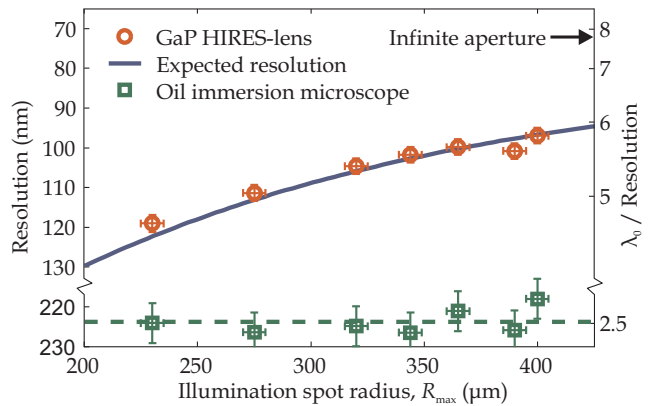}
  \caption{
  Optical resolution of a GaP HIRES-lens for different radii, $R_\text{max}$, of the illumination area.
  Red circles: measured resolutions of the HIRES-lens. Solid blue line: expected theoretical
  resolution deduced from Eq.~\ref{eq:intensityProfile}. Green squares: measured resolution of the oil immersion microscope.
  Dashed green line: mean measured resolution. Black arrow: expected resolution for an infinitely
  large illumination area. By increasing the illumination area the effective numerical aperture of the lens increases thereby improving the
  resolution.}\label{fig:Fig3}
\end{figure}
For a more quantitative study of the obtained resolution, we study the shape of
the focus in the HIRES-lens.
The radial intensity distribution of the focus is directly
calculated from a plane wave decomposition of the contributing
waves,
\begin{equation}
I(r) = I_0 \left[ k_\text{max}^2 \frac{J_1(k_\text{max} r)}{k_\text{max} r} -
k_\text{min}^2 \frac{J_1(k_\text{min} r)}{k_\text{min} r} \right]^2
\label{eq:intensityProfile}
\end{equation}
where $J_1$ is a Bessel function of the first kind.
The minimum and maximum coupled transversal k-vectors, $k_\text{min}$ and $k_\text{max}$,
are directly related to the inner and outer radius, $R_\text{min}$ and $R_\text{max}$,
of the illuminated area:
$k_\text{max} = n k_0 \left(1 + L^2/R_\text{max}^2 \right) ^{-\frac{1}{2}}$
(and similar for $k_\text{min}$).
To confirm this dependence, we imaged the objects for different
values of the illumination radius $R_\text{max}$. For each measurement the resolution is
determined by modeling the resulting image of a single $50$~nm gold
nanoparticle with Eq.~\ref{eq:intensityProfile}.
Since it is hard to quantify the resolution from the width of a non-Gaussian focal shape we
use Sparrow's criterion which defines the resolution as the minimal
distance at which two separate objects are still discernible, see e.g. \cite{Hecht1998aa}.
In Fig.~\ref{fig:Fig3} the measured resolution versus $R_\text{max}$ is shown.
As a reference we also plotted the measured resolution of the high-quality oil immersion microscope.
We see that the resolution improves as we increase the illuminated area. The measured resolutions are in
excellent correspondence with the expected
resolution obtained from the calculated intensity profile.
The resolution of the HIRES-lens is much better than the high-quality conventional oil immersion
microscope. The highest resolution we measured is $97 \pm 2$~nm, which demonstrates
imaging in the nanometer regime with visible wavelengths.

A GaP HIRES-lens has the potential to reach even better optical
resolutions up to $72$~nm. It is then possible to resolve
objects placed in each others near field at distances of $\lambda_0/2\pi$.
To achieve these resolutions
a wider area of the scattering porous layer has to be illuminated and as a result
light has to be scattered at increasingly higher angles from the
porous layer. Here advances could benefit from investigations in the
field of thin film solar cells where high angle scattering is beneficial
for optimal light harvesting\cite{Yablonovitch1982aa}.

Our results open the way to improve resolution in a wide range of
optical imaging techniques.
The robustness of a HIRES-lens against distortion and abberation,
together with their ease to manufacture, makes them ideal for the imaging of fluorescent
labeled biological samples or for the efficient coupling to
metamaterials\cite{Fang2005aa,Liu2007aa} and plasmonic
nanostructures\cite{Stockman2002aa,Aeschlimann2007aa,Kao2011aa}.
Recent developments in spatio-temporal control of waves in disordered
materials\cite{Aulbach2011aa,Katz2011aa,McCabe2011aa}
suggest the possibility for HIRES-lenses to create ultrashort pulses in
a nano-sized focus.
The fact that a HIRES-lens is a linear technique opens the possibility
to use it for resolution improvement of a large range of existing linear and non-linear
imaging techniques, such as confocal microscopy,
STED\cite{Dyba2002aa}, PALM\cite{Betzig2006aa}, and STORM\cite{Rust2006aa}.

\section{Acknowledgements}
The authors would like to acknowledge Hannie van den Broek, Cock Harteveld, L\'eon Woldering,
Willem Tjerkstra, Ivo Vellekoop, Christian Blum, and Vinod Subramaniam for their support
and insightful discussions.
This work is part of the research program
of the ``Stichting voor Fundamenteel Onderzoek der Materie (FOM)'', which is financially
supported by the ``Nederlandse Organisatie voor Wetenschappelijk Onderzoek (NWO)''.
JB is partially financed by the FIRB-MIUR "Futuro in Ricerca" project RBFR08UH60.
WLV thanks NWO-Vici and APM is supported by a Vidi grant from NWO.

\bibliographystyle{apsrev}
\bibliography{vanPutten2011}

\begin{thebibliography}{30}
\expandafter\ifx\csname natexlab\endcsname\relax\def\natexlab#1{#1}\fi
\expandafter\ifx\csname bibnamefont\endcsname\relax
  \def\bibnamefont#1{#1}\fi
\expandafter\ifx\csname bibfnamefont\endcsname\relax
  \def\bibfnamefont#1{#1}\fi
\expandafter\ifx\csname citenamefont\endcsname\relax
  \def\citenamefont#1{#1}\fi
\expandafter\ifx\csname url\endcsname\relax
  \def\url#1{\texttt{#1}}\fi
\expandafter\ifx\csname urlprefix\endcsname\relax\def\urlprefix{URL }\fi
\providecommand{\bibinfo}[2]{#2}
\providecommand{\eprint}[2][]{\url{#2}}

\bibitem[{\citenamefont{Hell and Wichmann}(1994)}]{Hell1994aa}
\bibinfo{author}{\bibfnamefont{S.~W.} \bibnamefont{Hell}} \bibnamefont{and}
  \bibinfo{author}{\bibfnamefont{J.}~\bibnamefont{Wichmann}},
  \bibinfo{journal}{Opt. Lett.} \textbf{\bibinfo{volume}{19}},
  \bibinfo{pages}{780} (\bibinfo{year}{1994}).

\bibitem[{\citenamefont{Dyba and Hell}(2002)}]{Dyba2002aa}
\bibinfo{author}{\bibfnamefont{M.}~\bibnamefont{Dyba}} \bibnamefont{and}
  \bibinfo{author}{\bibfnamefont{S.~W.} \bibnamefont{Hell}},
  \bibinfo{journal}{Phys. Rev. Lett.} \textbf{\bibinfo{volume}{88}},
  \bibinfo{pages}{163901} (\bibinfo{year}{2002}).

\bibitem[{\citenamefont{Betzig et~al.}(2006)\citenamefont{Betzig, Patterson,
  Sougrat, Lindwasser, Olenych, Bonifacino, Davidson, Lippincott-Schwartz, and
  Hess}}]{Betzig2006aa}
\bibinfo{author}{\bibfnamefont{E.}~\bibnamefont{Betzig}},
  \bibinfo{author}{\bibfnamefont{G.~H.} \bibnamefont{Patterson}},
  \bibinfo{author}{\bibfnamefont{R.}~\bibnamefont{Sougrat}},
  \bibinfo{author}{\bibfnamefont{O.~W.} \bibnamefont{Lindwasser}},
  \bibinfo{author}{\bibfnamefont{S.}~\bibnamefont{Olenych}},
  \bibinfo{author}{\bibfnamefont{J.~S.} \bibnamefont{Bonifacino}},
  \bibinfo{author}{\bibfnamefont{M.~W.} \bibnamefont{Davidson}},
  \bibinfo{author}{\bibfnamefont{J.}~\bibnamefont{Lippincott-Schwartz}},
  \bibnamefont{and} \bibinfo{author}{\bibfnamefont{H.~F.} \bibnamefont{Hess}},
  \bibinfo{journal}{Science} \textbf{\bibinfo{volume}{313}},
  \bibinfo{pages}{1642} (\bibinfo{year}{2006}).

\bibitem[{\citenamefont{Rust et~al.}(2006)\citenamefont{Rust, Bates, and
  Zhuang}}]{Rust2006aa}
\bibinfo{author}{\bibfnamefont{M.~J.} \bibnamefont{Rust}},
  \bibinfo{author}{\bibfnamefont{M.}~\bibnamefont{Bates}}, \bibnamefont{and}
  \bibinfo{author}{\bibfnamefont{X.}~\bibnamefont{Zhuang}},
  \bibinfo{journal}{Nat. Meth.} \textbf{\bibinfo{volume}{3}},
  \bibinfo{pages}{793} (\bibinfo{year}{2006}), ISSN \bibinfo{issn}{1548-7091}.

\bibitem[{\citenamefont{Hell}(2007)}]{Hell2007aa}
\bibinfo{author}{\bibfnamefont{S.~W.} \bibnamefont{Hell}},
  \bibinfo{journal}{Science} \textbf{\bibinfo{volume}{316}},
  \bibinfo{pages}{1153} (\bibinfo{year}{2007}).

\bibitem[{\citenamefont{Pohl and Courjon}(1993)}]{Pohl1993aa}
\bibinfo{author}{\bibfnamefont{D.}~\bibnamefont{Pohl}} \bibnamefont{and}
  \bibinfo{author}{\bibfnamefont{D.}~\bibnamefont{Courjon}},
  \emph{\bibinfo{title}{Near Field Optics}} (\bibinfo{publisher}{Kluwer,
  Dordrecht}, \bibinfo{year}{1993}).

\bibitem[{\citenamefont{Pendry}(2000)}]{Pendry2000aa}
\bibinfo{author}{\bibfnamefont{J.~B.} \bibnamefont{Pendry}},
  \bibinfo{journal}{Phys. Rev. Lett.} \textbf{\bibinfo{volume}{85}},
  \bibinfo{pages}{3966} (\bibinfo{year}{2000}).

\bibitem[{\citenamefont{Fang et~al.}(2005)\citenamefont{Fang, Lee, Sun, and
  Zhang}}]{Fang2005aa}
\bibinfo{author}{\bibfnamefont{N.}~\bibnamefont{Fang}},
  \bibinfo{author}{\bibfnamefont{H.}~\bibnamefont{Lee}},
  \bibinfo{author}{\bibfnamefont{C.}~\bibnamefont{Sun}}, \bibnamefont{and}
  \bibinfo{author}{\bibfnamefont{X.}~\bibnamefont{Zhang}},
  \bibinfo{journal}{Science} \textbf{\bibinfo{volume}{308}},
  \bibinfo{pages}{534} (\bibinfo{year}{2005}).

\bibitem[{\citenamefont{Liu et~al.}(2007)\citenamefont{Liu, Lee, Xiong, Sun,
  and Zhang}}]{Liu2007aa}
\bibinfo{author}{\bibfnamefont{Z.}~\bibnamefont{Liu}},
  \bibinfo{author}{\bibfnamefont{H.}~\bibnamefont{Lee}},
  \bibinfo{author}{\bibfnamefont{Y.}~\bibnamefont{Xiong}},
  \bibinfo{author}{\bibfnamefont{C.}~\bibnamefont{Sun}}, \bibnamefont{and}
  \bibinfo{author}{\bibfnamefont{X.}~\bibnamefont{Zhang}},
  \bibinfo{journal}{Science} \textbf{\bibinfo{volume}{315}},
  \bibinfo{pages}{1686} (\bibinfo{year}{2007}).

\bibitem[{\citenamefont{Stockman et~al.}(2002)\citenamefont{Stockman, Faleev,
  and Bergman}}]{Stockman2002aa}
\bibinfo{author}{\bibfnamefont{M.~I.} \bibnamefont{Stockman}},
  \bibinfo{author}{\bibfnamefont{S.~V.} \bibnamefont{Faleev}},
  \bibnamefont{and} \bibinfo{author}{\bibfnamefont{D.~J.}
  \bibnamefont{Bergman}}, \bibinfo{journal}{Phys. Rev. Lett.}
  \textbf{\bibinfo{volume}{88}}, \bibinfo{pages}{067402}
  (\bibinfo{year}{2002}).

\bibitem[{\citenamefont{Aeschlimann et~al.}(2007)\citenamefont{Aeschlimann,
  Bauer, Bayer, Brixner, Garcia~de Abajo, Pfeiffer, Rohmer, Spindler, and
  Steeb}}]{Aeschlimann2007aa}
\bibinfo{author}{\bibfnamefont{M.}~\bibnamefont{Aeschlimann}},
  \bibinfo{author}{\bibfnamefont{M.}~\bibnamefont{Bauer}},
  \bibinfo{author}{\bibfnamefont{D.}~\bibnamefont{Bayer}},
  \bibinfo{author}{\bibfnamefont{T.}~\bibnamefont{Brixner}},
  \bibinfo{author}{\bibfnamefont{F.~J.} \bibnamefont{Garcia~de Abajo}},
  \bibinfo{author}{\bibfnamefont{W.}~\bibnamefont{Pfeiffer}},
  \bibinfo{author}{\bibfnamefont{M.}~\bibnamefont{Rohmer}},
  \bibinfo{author}{\bibfnamefont{C.}~\bibnamefont{Spindler}}, \bibnamefont{and}
  \bibinfo{author}{\bibfnamefont{F.}~\bibnamefont{Steeb}},
  \bibinfo{journal}{Nature} \textbf{\bibinfo{volume}{446}},
  \bibinfo{pages}{301} (\bibinfo{year}{2007}), ISSN \bibinfo{issn}{0028-0836}.

\bibitem[{\citenamefont{Kao et~al.}(2011)\citenamefont{Kao, Jenkins,
  Ruostekoski, and Zheludev}}]{Kao2011aa}
\bibinfo{author}{\bibfnamefont{T.~S.} \bibnamefont{Kao}},
  \bibinfo{author}{\bibfnamefont{S.~D.} \bibnamefont{Jenkins}},
  \bibinfo{author}{\bibfnamefont{J.}~\bibnamefont{Ruostekoski}},
  \bibnamefont{and} \bibinfo{author}{\bibfnamefont{N.~I.}
  \bibnamefont{Zheludev}}, \bibinfo{journal}{Phys. Rev. Lett.}
  \textbf{\bibinfo{volume}{106}}, \bibinfo{pages}{085501}
  (\bibinfo{year}{2011}).

\bibitem[{\citenamefont{Vellekoop and Mosk}(2007)}]{Vellekoop2007aa}
\bibinfo{author}{\bibfnamefont{I.~M.} \bibnamefont{Vellekoop}}
  \bibnamefont{and} \bibinfo{author}{\bibfnamefont{A.~P.} \bibnamefont{Mosk}},
  \bibinfo{journal}{Opt. Lett.} \textbf{\bibinfo{volume}{32}},
  \bibinfo{pages}{2309} (\bibinfo{year}{2007}).

\bibitem[{\citenamefont{Leith and Upatnieks}(1966)}]{Leith1966aa}
\bibinfo{author}{\bibfnamefont{E.~N.} \bibnamefont{Leith}} \bibnamefont{and}
  \bibinfo{author}{\bibfnamefont{J.}~\bibnamefont{Upatnieks}},
  \bibinfo{journal}{J. Opt. Soc. Am.} \textbf{\bibinfo{volume}{56}},
  \bibinfo{pages}{523} (\bibinfo{year}{1966}).

\bibitem[{\citenamefont{Fink et~al.}(2000)\citenamefont{Fink, Cassereau,
  Derode, Prada, Roux, Tanter, Thomas, and Wu}}]{Fink2000}
\bibinfo{author}{\bibfnamefont{M.}~\bibnamefont{Fink}},
  \bibinfo{author}{\bibfnamefont{D.}~\bibnamefont{Cassereau}},
  \bibinfo{author}{\bibfnamefont{A.}~\bibnamefont{Derode}},
  \bibinfo{author}{\bibfnamefont{C.}~\bibnamefont{Prada}},
  \bibinfo{author}{\bibfnamefont{P.}~\bibnamefont{Roux}},
  \bibinfo{author}{\bibfnamefont{M.}~\bibnamefont{Tanter}},
  \bibinfo{author}{\bibfnamefont{J.-L.} \bibnamefont{Thomas}},
  \bibnamefont{and} \bibinfo{author}{\bibfnamefont{F.}~\bibnamefont{Wu}},
  \bibinfo{journal}{Rep. Prog. Phys.} \textbf{\bibinfo{volume}{63}},
  \bibinfo{pages}{1933–} (\bibinfo{year}{2000}).

\bibitem[{\citenamefont{Wu et~al.}(1999)\citenamefont{Wu, Feke, Grober, and
  Ghislain}}]{Wu1999aa}
\bibinfo{author}{\bibfnamefont{Q.}~\bibnamefont{Wu}},
  \bibinfo{author}{\bibfnamefont{G.~D.} \bibnamefont{Feke}},
  \bibinfo{author}{\bibfnamefont{R.~D.} \bibnamefont{Grober}},
  \bibnamefont{and} \bibinfo{author}{\bibfnamefont{L.~P.}
  \bibnamefont{Ghislain}}, \bibinfo{journal}{Appl. Phys. Lett.}
  \textbf{\bibinfo{volume}{75}}, \bibinfo{pages}{4064} (\bibinfo{year}{1999}).

\bibitem[{\citenamefont{Axelrod et~al.}(1984)\citenamefont{Axelrod, Burghardt,
  and Thompson}}]{Axelrod1984aa}
\bibinfo{author}{\bibfnamefont{D.}~\bibnamefont{Axelrod}},
  \bibinfo{author}{\bibfnamefont{T.~P.} \bibnamefont{Burghardt}},
  \bibnamefont{and} \bibinfo{author}{\bibfnamefont{N.~L.}
  \bibnamefont{Thompson}}, \bibinfo{journal}{Annu. Rev. Biophys. Bioeng.}
  \textbf{\bibinfo{volume}{13}}, \bibinfo{pages}{247–268}
  (\bibinfo{year}{1984}).

\bibitem[{\citenamefont{Vellekoop et~al.}(2010)\citenamefont{Vellekoop,
  Lagendijk, and Mosk}}]{Vellekoop2010aa}
\bibinfo{author}{\bibfnamefont{I.}~\bibnamefont{Vellekoop}},
  \bibinfo{author}{\bibfnamefont{A.}~\bibnamefont{Lagendijk}},
  \bibnamefont{and} \bibinfo{author}{\bibfnamefont{A.}~\bibnamefont{Mosk}},
  \bibinfo{journal}{Nat Photon} \textbf{\bibinfo{volume}{4}},
  \bibinfo{pages}{320} (\bibinfo{year}{2010}).

\bibitem[{\citenamefont{Feng et~al.}(1988)\citenamefont{Feng, Kane, Lee, and
  Stone}}]{Feng1988aa}
\bibinfo{author}{\bibfnamefont{S.}~\bibnamefont{Feng}},
  \bibinfo{author}{\bibfnamefont{C.}~\bibnamefont{Kane}},
  \bibinfo{author}{\bibfnamefont{P.~A.} \bibnamefont{Lee}}, \bibnamefont{and}
  \bibinfo{author}{\bibfnamefont{A.~D.} \bibnamefont{Stone}},
  \bibinfo{journal}{Phys. Rev. Lett.} \textbf{\bibinfo{volume}{61}},
  \bibinfo{pages}{834} (\bibinfo{year}{1988}).

\bibitem[{\citenamefont{Freund et~al.}(1988)\citenamefont{Freund, Rosenbluh,
  and Feng}}]{Freund1988aa}
\bibinfo{author}{\bibfnamefont{I.}~\bibnamefont{Freund}},
  \bibinfo{author}{\bibfnamefont{M.}~\bibnamefont{Rosenbluh}},
  \bibnamefont{and} \bibinfo{author}{\bibfnamefont{S.}~\bibnamefont{Feng}},
  \bibinfo{journal}{Phys. Rev. Lett.} \textbf{\bibinfo{volume}{61}},
  \bibinfo{pages}{2328} (\bibinfo{year}{1988}).

\bibitem[{\citenamefont{Vellekoop and Aegerter}(2010)}]{Vellekoop2010ab}
\bibinfo{author}{\bibfnamefont{I.}~\bibnamefont{Vellekoop}} \bibnamefont{and}
  \bibinfo{author}{\bibfnamefont{C.}~\bibnamefont{Aegerter}},
  \bibinfo{journal}{Opt. Lett.} \textbf{\bibinfo{volume}{35}},
  \bibinfo{pages}{1245} (\bibinfo{year}{2010}).

\bibitem[{\citenamefont{Aspnes and Studna}(1983)}]{Aspnes1983aa}
\bibinfo{author}{\bibfnamefont{D.~E.} \bibnamefont{Aspnes}} \bibnamefont{and}
  \bibinfo{author}{\bibfnamefont{A.~A.} \bibnamefont{Studna}},
  \bibinfo{journal}{Phys. Rev. B} \textbf{\bibinfo{volume}{27}},
  \bibinfo{pages}{985} (\bibinfo{year}{1983}).

\bibitem[{\citenamefont{Schuurmans et~al.}(1999)\citenamefont{Schuurmans,
  Vanmaekelbergh, van~de Lagemaat, and Lagendijk}}]{Schuurmans1999aa}
\bibinfo{author}{\bibfnamefont{F.~J.~P.} \bibnamefont{Schuurmans}},
  \bibinfo{author}{\bibfnamefont{D.}~\bibnamefont{Vanmaekelbergh}},
  \bibinfo{author}{\bibfnamefont{J.}~\bibnamefont{van~de Lagemaat}},
  \bibnamefont{and}
  \bibinfo{author}{\bibfnamefont{A.}~\bibnamefont{Lagendijk}},
  \bibinfo{journal}{Science} \textbf{\bibinfo{volume}{284}},
  \bibinfo{pages}{141} (\bibinfo{year}{1999}).

\bibitem[{Sup(2011)}]{SupportingMaterial_vanPutten2011aa}
\emph{\bibinfo{title}{Details on materials and methods are forthcoming.}}
  (\bibinfo{year}{2011}).

\bibitem[{\citenamefont{Hsieh et~al.}(2010)\citenamefont{Hsieh, Pu, Grange,
  Laporte, and Psaltis}}]{Hsieh2010aa}
\bibinfo{author}{\bibfnamefont{C.-L.} \bibnamefont{Hsieh}},
  \bibinfo{author}{\bibfnamefont{Y.}~\bibnamefont{Pu}},
  \bibinfo{author}{\bibfnamefont{R.}~\bibnamefont{Grange}},
  \bibinfo{author}{\bibfnamefont{G.}~\bibnamefont{Laporte}}, \bibnamefont{and}
  \bibinfo{author}{\bibfnamefont{D.}~\bibnamefont{Psaltis}},
  \bibinfo{journal}{Opt. Express} \textbf{\bibinfo{volume}{18}},
  \bibinfo{pages}{20723} (\bibinfo{year}{2010}).

\bibitem[{\citenamefont{Hecht}(1998)}]{Hecht1998aa}
\bibinfo{author}{\bibfnamefont{E.}~\bibnamefont{Hecht}},
  \emph{\bibinfo{title}{Optics}} (\bibinfo{publisher}{Addison Wesley Longman,
  Inc.}, \bibinfo{year}{1998}).

\bibitem[{\citenamefont{Yablonovitch and Cody}(1982)}]{Yablonovitch1982aa}
\bibinfo{author}{\bibfnamefont{E.}~\bibnamefont{Yablonovitch}}
  \bibnamefont{and} \bibinfo{author}{\bibfnamefont{G.~D.} \bibnamefont{Cody}},
  \bibinfo{journal}{IEEE Trans. Electron Devices}
  \textbf{\bibinfo{volume}{29}}, \bibinfo{pages}{300–} (\bibinfo{year}{1982}).

\bibitem[{\citenamefont{Aulbach et~al.}(2011)\citenamefont{Aulbach, Gjonaj,
  Johnson, Mosk, and Lagendijk}}]{Aulbach2011aa}
\bibinfo{author}{\bibfnamefont{J.}~\bibnamefont{Aulbach}},
  \bibinfo{author}{\bibfnamefont{B.}~\bibnamefont{Gjonaj}},
  \bibinfo{author}{\bibfnamefont{P.~M.} \bibnamefont{Johnson}},
  \bibinfo{author}{\bibfnamefont{A.~P.} \bibnamefont{Mosk}}, \bibnamefont{and}
  \bibinfo{author}{\bibfnamefont{A.}~\bibnamefont{Lagendijk}},
  \bibinfo{journal}{Phys. Rev. Lett.} \textbf{\bibinfo{volume}{106}},
  \bibinfo{pages}{103901} (\bibinfo{year}{2011}).

\bibitem[{\citenamefont{Katz et~al.}(2011)\citenamefont{Katz, Bromberg, Small,
  and Silberberg}}]{Katz2011aa}
\bibinfo{author}{\bibfnamefont{O.}~\bibnamefont{Katz}},
  \bibinfo{author}{\bibfnamefont{Y.}~\bibnamefont{Bromberg}},
  \bibinfo{author}{\bibfnamefont{E.}~\bibnamefont{Small}}, \bibnamefont{and}
  \bibinfo{author}{\bibfnamefont{Y.}~\bibnamefont{Silberberg}},
  \bibinfo{journal}{arXiv:1012.0413}  (\bibinfo{year}{2011}).

\bibitem[{\citenamefont{McCabe et~al.}(2011)\citenamefont{McCabe, Tajalli,
  Austin, Bondareff, Walmsley, Gigan, and Chatel}}]{McCabe2011aa}
\bibinfo{author}{\bibfnamefont{D.~J.} \bibnamefont{McCabe}},
  \bibinfo{author}{\bibfnamefont{A.}~\bibnamefont{Tajalli}},
  \bibinfo{author}{\bibfnamefont{D.~R.} \bibnamefont{Austin}},
  \bibinfo{author}{\bibfnamefont{P.}~\bibnamefont{Bondareff}},
  \bibinfo{author}{\bibfnamefont{I.~A.} \bibnamefont{Walmsley}},
  \bibinfo{author}{\bibfnamefont{S.}~\bibnamefont{Gigan}}, \bibnamefont{and}
  \bibinfo{author}{\bibfnamefont{B.}~\bibnamefont{Chatel}},
  \bibinfo{journal}{arXiv:1101.0976}  (\bibinfo{year}{2011}).

\end{thebibliography}
\end{document}